\documentclass[preprint,aps,nofootinbib]{revtex4}
\input{epsf.tex}
\usepackage[dvips]{graphicx,epsfig}
\usepackage{subfigure}
\usepackage{textcomp}
\def\bkR{{\rm I\kern-.17em R}}
\def\bkC{{\rm \kern.24em \vrule width.05em height1.4ex depth-.05ex \kern-.26em C}}

\def\be{\beta}

\def\frac#1#2{{\textstyle{{#1}\over {#2}}}}

\def\lsim{\mathrel{\rlap{\lower4pt\hbox{\hskip1pt$\sim$}}
    \raise1pt\hbox{$<$}}}
\def\gsim{\mathrel{\rlap{\lower4pt\hbox{\hskip1pt$\sim$}}
    \raise1pt\hbox{$>$}}}
\def\sqr#1#2{{\vcenter{\vbox{\hrule height.#2pt
         \hbox{\vrule width.#2pt height#1pt \kern#1pt
         \vrule width.#2pt}
         \hrule height.#2pt}}}}

\def\laq{\raise 0.4 ex \hbox{$<$}\kern -0.8 em\lower 0.62 ex\hbox{$\sim$}}
\def\gaq{\raise 0.4 ex \hbox{$>$}\kern -0.7 em\lower 0.62 ex\hbox{$\sim$}}

\def\be{\begin{equation}}
\def\ee{\end{equation}}
\def\ba{\begin{eqnarray}}
\def\ea{\end{eqnarray}}

\def\dalemb#1#2{{\vbox{\hrule height.#2pt
        \hbox{\vrule width.#2pt height#1pt \kern#1pt \vrule width.#2pt}
        \hrule height.#2pt}}}

\def\dalemb#1#2{{\vbox{\hrule height.#2pt
        \hbox{\vrule width.#2pt height#1pt \kern#1pt \vrule width.#2pt}
        \hrule height.#2pt}}}

\def\gtorder{\mathrel{\raise.3ex\hbox{$>$}\mkern-14mu
             \lower0.6ex\hbox{$\sim$}}}
\def\ltorder{\mathrel{\raise.3ex\hbox{$<$}\mkern-14mu
             \lower0.6ex\hbox{$\sim$}}}

\begin{document}

\rightline{July 2012}

\title{Noncommutative Graphene}

\author{Catarina Bastos\footnote{E-mail: catarina.bastos@ist.utl.pt}}

\vskip 0.3cm

\affiliation{Instituto de Plasmas e Fus\~ao Nuclear, Instituto Superior T\'ecnico \\
Avenida Rovisco Pais 1, 1049-001 Lisboa, Portugal}

\author{Orfeu Bertolami\footnote{Also at Instituto de Plasmas e Fus\~ao Nuclear,
Instituto Superior T\'ecnico. E-mail: orfeu.bertolami@fc.up.pt}}

\vskip 0.3cm

\affiliation{Departamento de F\'\i sica e Astronomia, \\
Faculdade de Ci\^encias da Universidade do Porto \\
Rua do Campo Alegre, 687,4169-007 Porto, Portugal}

\author{Nuno Costa Dias\footnote{Also at Grupo de F\'{\i}sica Matem\'atica, UL,
Avenida Prof. Gama Pinto 2, 1649-003, Lisboa, Portugal. E-mail: ncdias@meo.pt}, Jo\~ao Nuno Prata\footnote{Also at Grupo de F\'{\i}sica Matem\'atica, UL,
Avenida Prof. Gama Pinto 2, 1649-003, Lisboa, Portugal. E-mail: joao.prata@mail.telepac.pt}}

\vskip 0.3cm

\affiliation{Departamento de Matem\'{a}tica, Universidade Lus\'ofona de
Humanidades e Tecnologias \\
Avenida Campo Grande, 376, 1749-024 Lisboa, Portugal}


\vskip 2cm

\begin{abstract}
We consider a noncommutative description of graphene. This
description consists of a Dirac equation for massless Dirac
fermions plus noncommutative corrections, which are treated in the
presence of an external magnetic field. We argue that, being a
two-dimensional Dirac system, graphene is particularly interesting
to test noncommutativity. We find that momentum noncommutativity
affects the energy levels of graphene, but that it does not entail
any kind of correction to the Hall conductivity.
\end{abstract}

\maketitle

\section{Introduction}

Graphene is a two-dimensional configuration of carbon atoms
organized in a hexagonal honeycomb structure
\cite{Novoselov1,Neto,Peres}. Often, a crystal lattice is a
Bavrais lattice, that is, an infinite array of discrete points
with an organization and orientation that appears exactly the
same, from whatever point the array is viewed. However, the
graphene hexagonal lattice is non-Bravais, as only the
next-to-nearest neighbor points appear with the same organization
and orientation. In the case of graphene one has two triangular
Bravais lattices, $A$ and $B$ which together form the non-Bravais
graphene lattice, and the difference between them is a rotation of
$\pi$. The hexagonal lattice belongs to the class of bipartite
lattices, and so one can say that graphene is a bipartite
non-Bravais lattice with two carbon atoms per unit cell. For the
two sub-lattices one has the same primitive vectors, which depend
explicitly on the distance between the two lattice points. With
these primitive vectors one can characterize any space point as a
linear combination of them. In the momentum space, one can obtain
the reciprocal primitive vectors and the corresponding Brillouin
zones, i.e. a uniquely defined primitive cell in the reciprocal
space. The first Brillouin zone forms a hexagon, which is rotated
by $\pi/12$ compared to the hexagonal structure in position space
and the corners of the first Brillouin zone are usually organized
in two sets, the Dirac points $K$ and $K'$. They are six Dirac
points in the total, but only two are worth considering due to the
periodicity of the momenta in the Brillouin zone \cite{Neto}.

It turns out that graphene's low energy excitations are
relativistic corresponding to massless, quasi-free fermions that
can be theoretically described by the Dirac equation for these
particles \cite{Peres}. Thus, one considers the Dirac equation at
the vicinity of the Dirac points $K$ and $K'$. Expanding the
dispersion relation around these points, one has to a first order
approximation a linear relation, which, for small energies, gives
origin to the so-called Dirac cones. These cones imply that
graphene can be seen as a conventional semiconductor, given that
there is no gap between conduction and valence bands.

We start with Dirac equation,
 \be\label{eq1}
i\hbar{\partial\psi\over{\partial t}}= H_D\psi~,
\ee 
where the wave function $\psi$ in the graphene case describes the electron
states around the Dirac points $K$ and $K'$, and the Dirac
Hamiltonian is given by \cite{Geim} 
\be\label{eq2}
H_D=c[\vec{\alpha}\cdot\vec{P}+{\bf\beta}mc]~, 
\ee 
where $\vec{\alpha}$ and ${\bf\beta}$ are the Dirac matrices and
$\vec{P}=\left(-i\hbar{\partial_x}, -i\hbar{\partial_y},0
\right)$. In the case of graphene, one has massless particles that
move through the honeycomb lattice with a velocity $v_{F}\sim
10^6~ms^{-1}$, the so-called Fermi velocity. Thus, for instance,
the Dirac Hamiltonian around the Dirac point $K$ reads
\be\label{eq3}
H=v_{F}\vec{\sigma}\cdot \vec{P}~. 
\ee 
where, $\vec{\sigma}=(\sigma_x,\sigma_y,\sigma_z)$ and $\sigma_x $, $\sigma_y$ and $\sigma_z$ are the Pauli matrices. The same Hamiltonian can be written to the Dirac Point $K'$ with  $\vec{\sigma}^{*}$ given by $\vec{\sigma}^{*}=(\sigma_x,-\sigma_y,\sigma_z)$. We can now write the
Hamiltonian for the two Dirac points as \cite{Geim},
\begin{equation}\label{eq4}
H_D=\left( \begin{array}{cc}
H_K & 0\\
0 & H_{K'}\\
\end{array}\right)=v_F\left( \begin{array}{cccc}
0 & p_x-ip_y & 0 & 0\\
p_x+ip_y & 0 & 0 & 0\\
0 & 0 & 0 & p_x+ip_y\\
0 & 0 & p_x-ip_y & 0\\
\end{array}\right)~.
\end{equation}
As mentioned, the wave function $\psi$ consists of two components,
one describing $K$ and other $K'$. Moreover, for these two Dirac
points one has an eigenvector that describes the probablility of
an electron state to be on sub-lattice A in the upper component,
or on the sub-lattice B in the lower component of the eigenstate.
Thus,
\begin{equation}\label{eq5}
\psi=\left( \begin{array}{c}
\psi^K\\
\psi^{K'}\\
\end{array}\right)~,
\end{equation}
where $\psi^K$ and $\psi^{K'}$ are two dimensional eigenstantes,
\begin{equation}\label{eq5a}
\psi^K=\left( \begin{array}{c}
\phi^{A}\\
\phi^{B}\\
\end{array}\right)\hspace{0,2cm},\hspace{0,2cm}
\psi^{K'}=\left( \begin{array}{c}
\phi^{A'}\\
\phi^{B'}\\
\end{array}\right)
\end{equation}
To obtain the dispersion relation for the energy one has to solve the following eigenvalue problem for each Dirac point,
\ba\label{eq6}
H_K\psi^K&=&E_K\psi^K~,\nonumber\\
H_{K'}\psi^{K'}&=&E_{K'}\psi^{K'}~.
 \ea
We have for the eigenvalues, for the two Dirac points $K$ and $K'$,
\cite{Geim}
\be\label{eq7}
 E_{K,K'}=\pm \hbar v_{F} |\vec{k}|
 \ee and to
evaluate the eigenvectors of the system we consider Eq. (\ref{eq6})
in the momentum representation, 
\ba\label{eq8} 
(\vec{\sigma}\cdot\hat {k})\psi^K(\vec{k})&=&\pm \psi^K(\vec{k})\\
(\vec{\sigma}^{*}\cdot\hat {k})\psi^{K'}(\vec{k})&=&\pm \psi^{K'}(\vec{k})~,
\ea 
where $\vec{\sigma}$ and $\vec{\sigma}^{*}$ are defined above. Moreover, these eigenvectors in the momentum space, for the two Dirac points, have the form \cite{Neto}
 \begin{equation}\label{eq8a}
\psi^K(\vec{k})={1\over\sqrt{2}}\left( \begin{array}{c}
e^{-i\varphi_{k}/2}\\
\pm e^{i\varphi_{k}/2}\\
\end{array}\right)\hspace{0,2cm},\hspace{0,2cm}
\psi^{K'}={1\over\sqrt{2}}\left( \begin{array}{c}
e^{i\varphi_{k}/2}\\
\pm e^{-i\varphi_{k}/2}\\
\end{array}\right)~,
\end{equation}
where $\varphi_k$ is the polar angle of vector $\vec{k}$ in the momentum space. The signs $\pm$ correspond to the eigenvalues of the energy spectrum for each Dirac point, given by Eq. (\ref{eq7}). 
One clearly sees
that the pseudo spinor $\psi(\vec{k})$ has a definite
pseudo-helicity. Furthermore, the existence of chiral symmetry, in
the previous eigenvectors, allows for the observation of the
anomalous quantum Hall effect
\cite{Peres,Geim,Gusynin,Novoselov2}. This is a striking feature
of graphene, which instead of the usual Landau levels for a
semiconductor in a magnetic field, it exhibits a different
degeneracy factor. This is because the fundamental energy level of
the graphene has two valleys \cite{Gusynin,Novoselov2}.

It has been argued in various instances that noncommutativity
should be considered as a fundamental feature of space-time at the
Planck scale \cite{CS}. Various noncommutative field theory models
\cite{DS,OBLG,OBCZ,Queiroz} have been discussed as well as many
extensions of quantum mechanics
\cite{Nair,Gamboa,DK,Horwathy,ZA,OB,OB2,Bastos1,Dias,Bastos9}. Of
particular interest is the so-called phase-space noncommutativity
which has been investigated in the context of quantum cosmology
\cite{Bastos2}, black holes physics and the singularity problem
\cite{Bastos3,Bastos4,Bastos5}. The phase-space noncommutative
algebra is given by \cite{Bastos1}
\begin{equation}
\left[\hat x_i, \hat x_j \right] = i \theta_{ij} , \hspace{0.5 cm} \left[\hat x_i, \hat p_j \right] = i \hbar \delta_{ij} ,
\hspace{0.5 cm} \left[\hat p_i, \hat p_j \right] = i \eta_{ij} ,  \hspace{0.5 cm} i,j= 1, ... ,d
\label{Eq0.1}
\end{equation}
where $\eta_{ij}$ and $\theta_{ij}$ are antisymmetric real constant ($d \times d$) matrices and $\delta_{ij}$ is the identity matrix. The key property of this extended algebra is that it is related to the standard Heisenberg-Weyl algebra:
\begin{equation}
\left[\hat x'_i, \hat x'_j \right] = 0 , \hspace{0.5 cm} \left[\hat x'_i, \hat p'_j \right]
= i \hbar \delta_{ij} , \hspace{0.5 cm} \left[\hat p'_i, \hat p'_j \right] = 0 ,
\hspace{0.5 cm} i,j= 1, ... ,d ~,
\label{Eq0.2}
\end{equation}
by a class of linear (non-canonical) transformations:
\begin{equation}
\hat x_i = \hat x_i \left(\hat x'_j , \hat p'_j \right) \hspace{1 cm}
\hat p_i = \hat p_i \left(\hat x'_j , \hat p'_j \right)~.
\label{Eq0.3}
\end{equation}

More recently, a new representation of noncommutativity has been
proposed \cite{Falomir1}. This representation uses the Pauli
matrices as the fundamental elements of the noncommutative
algebra. It allows for a noncommutative extension of graphene by
treating it as a quantum charged particle subjected to an
electromagnetic field \cite{Falomir2}.

Since graphene is a two-dimensional system it is expected to
supply an interesting model where to test noncommutativity. First,
because it is a special system where quantum relativistic
phenomena, typical of high-energy physics, arise at low-energies.
Second because in non-relativistic quantum mechanics time is
always a commutative variable and so noncommutativity, in this
context, can only be applied to the spatial variables. Since any
sympletic form in an odd dimension is always degenerate, after a
linear transformation we can only have noncommutativity in two
dimensions. Graphene provides a real two-dimensional physical
system. Various bounds and inequalities, which appear in the
context of two-dimensional noncommutative quantum mechanics, where
derived in \cite{Bastos1}.

In this work, we extend the model of graphene to a noncommutative
phase-space setting. We consider graphene in an external magnetic
field and a noncommutative geometry. In section II A we review the
concepts behind the graphene model in the presence of a magnetic
field. Then, in section II B we present the noncommutative
extension of the graphene in a constant magnetic field. In section
II C we examine the effect of noncommutativity on graphene's
anomalous quantum Hall effect. In section III we compare the
experimental values obtained for the energy levels of graphene
with our theoretical predictions and obtain a bound for the
noncommutative parameter $\eta$. Finally, in section IV, we
summarize the main conclusions and results.

\section{Graphene in an external magnetic field}

In this section, we start by obtaining the energy dispersion
relation for a layer of graphene subjected to an external constant
magnetic field. We then show how to extend the problem to a
phase-space noncommutative setting, and consider the implications
of this extension on the anomalous quantum Hall effect.

\subsection{The Commutative Case}

Let us consider a layer of graphene in a external constant
magnetic field, $ \vec{B}=B \vec{e}_z$. In our units $c=1$. We
introduce the $\vec{B}$-field through the minimal coupling to the
vector potential, such that $\vec{B}=\vec{\nabla}\times\vec{A}$
\be\label{eq1.1} \vec{P}\rightarrow \vec{P}-{e}\vec{A} ~, \ee
where \be\label{eq1.2} \vec{A}= {B \over 2}\left(-y,x,0\right)~,
\ee and $e$ is the charge of the electron. Thus, for the two Dirac
points $K$ and $K'$ the Hamiltonians read \cite{Neto}:
\be\label{eq1.3} H_K=v_F\left( \begin{array}{cc}
0 & p_x-ip_y +{eB\over2}(y+ix)\\
p_x+ip_y +{eB\over2}(y-ix) & 0
\end{array}\right)~,
\ee
\be\label{eq1.4}
H_{K'}=v_F\left( \begin{array}{cc}
 0 &p_x+ip_y +{eB\over2}(y-ix)\\
p_x-ip_y +{eB\over2}(y+ix) & 0\\
\end{array}\right)
\ee and the energy eigenvalue equation for the wave function Eq.
(\ref{eq5}), at the Dirac point $K$, is then:
\ba\label{eq1.5I}
{\hbar v_F\over \xi}\left[-i\left(\xi\partial_x-{x\over \xi}\right)+\left(- \xi\partial_y+{y\over \xi}\right)\right]\phi^B=E_K\phi^A~,\nonumber\\
{\hbar v_F\over \xi}\left[-i\left(\xi\partial_x+{x\over \xi}\right)+\left(\xi\partial_y+{y\over \xi}\right)\right]\phi^A=E_K\phi^B~,
\ea
where $\xi=\sqrt{{2\hbar\over{eB}}}=\sqrt{2}l_B$ is an auxiliary variable and $l_B$ is the so-called magnetic length \cite{Neto}.

Redefining the variables as,
\ba\label{eq1.6}
\partial_{\tilde{x}}=\xi\partial_x\hspace{0,25cm}&,&\hspace{0,25cm}\partial_{\tilde{y}}=\xi\partial_y~,\nonumber\\
\tilde{x}={x\over
\xi}\hspace{0,25cm}&,&\hspace{0,25cm}\tilde{y}={y\over \xi}~, \ea
the system, Eq. (\ref{eq1.5I}), turns into
\ba\label{eq1.5}
&&{\hbar v_F\over \xi}\left[-i\left(\partial_{\tilde{x}}-\tilde{x}\right)+\left(-\partial_{\tilde{y}}+\tilde{y}\right)\right]\phi^B=E_K\phi^A~,\nonumber \hspace{2cm}(a)\\
&&{\hbar v_F\over\xi}\left[-i\left(\partial_{\tilde{x}}+\tilde{x}\right)+\left(\partial_{\tilde{y}}+\tilde{y}\right)\right]\phi^A=E_K\phi^B~.
\hspace{2,4cm} (b) \ea Solving Eq. (\ref{eq1.5}b) for $\phi^B$ and
substituting into Eq. (\ref{eq1.5}a), yields \be\label{eq1.6}
\left({\hbar v_F\over \xi}\right)^2
\left[-i\left(\partial_{\tilde{x}}-\tilde{x}\right)+\left(-\partial_{\tilde{y}}+\tilde{y}\right)\right]
\left[-i\left(\partial_{\tilde{x}}+\tilde{x}\right)+\left(\partial_{\tilde{y}}+\tilde{y}\right)\right]\phi^A=E_K^2\phi^A~,
\ee which is a second order equation analogous to the energy
eigenvalue equation for the quantum harmonic oscillator in two
dimensions. It can be solved using the set of annihilation and
creation operators, \ba\label{eq1.7}
a_x={1\over\sqrt{2}}(\tilde{x}+\partial_{\tilde{x}})\hspace{0,2cm},\hspace{0,2cm}a_x^{\dag}={1\over\sqrt{2}}(\tilde{x}-\partial_{\tilde{x}})~,\nonumber\\
a_y={1\over\sqrt{2}}(\tilde{y}+\partial_{\tilde{y}})\hspace{0,2cm},\hspace{0,2cm}a_y^{\dag}={1\over\sqrt{2}}(\tilde{y}-\partial_{\tilde{y}})~.
\ea
However, it is more convenient to use left/ right operators, as in the case of the quantum harmonic oscillator in two dimensions. These new operators are responsible not only for adding (subtracting) a quantum of energy, but also for adding (subtracting) a quantum of angular momentum, $\hbar$, in the direct direction (right operators with subscript $d$) or in the inverse direction (left operators with subscript $e$). Furthermore they simplify considerably the calculations.Thus, one introduces the left/ right operators as,
\ba\label{eq1.8}
a_d={1\over\sqrt{2}}(a_x-ia_y)\hspace{0,2cm},\hspace{0,2cm}a_d^{\dag}={1\over\sqrt{2}}(a_x^{\dag}+ia_y^{\dag})~,\nonumber\\
a_e={1\over\sqrt{2}}(a_x+ia_y)\hspace{0,2cm},\hspace{0,2cm}a_e^{\dag}={1\over\sqrt{2}}(a_x^{\dag}-ia_y^{\dag})~.
\ea
So, Eq. (\ref{eq1.6}) can now be rewritten in terms of these left / right operators,
\be\label{eq1.9}
4\left({\hbar v_F\over \xi}\right)^2 (a_e^{\dag} a_e )  \phi^A=E_K^2 \phi^A~.
\ee
One sees that the Hamiltonian depends only on the left operators. Note that for the angular momentum one has $l_z=\hbar(n_d-n_e)\equiv\hbar m$, where $n_e$ and $n_d$ are the eigenvalues associated to the left and right number operators $(N_e=a_e^{\dag} a_e,~N_d=a_d^{\dag} a_d)$ respectively; $m$ is the eigenvalue associated to the angular momentum operator in the $z$-direction. Let us suppose that $\phi^A=\phi_{(n_e,n_d)}^A$, then the energy spectrum for the Dirac point $K$ is given by
\be\label{eq1.10}
E_K=\pm2{\hbar v_F\over \xi}\sqrt{n_e}=\pm {\sqrt{2}\hbar v_F\over l_B}\sqrt{n_e}~,
\ee
with $n_e=0,1,2,...$.

Finally, if we consider the eigenvalue problem, Eq.
(\ref{eq1.5}b), and substitute the egeinvalue $E_K$ obtained in
Eq. (\ref{eq1.10}), we get the eigenstates for the Dirac $K$
point.
\be\label{eq1.11} \psi_{(n_e,n_d)}^K=\left(
\begin{array}{c}
\phi_{(n_e,n_d)}^{A}\\
\pm i\phi_{(n_e-1,n_d)}^{B}\\
\end{array}\right)\hspace{0,2cm},\hspace{0,2cm}
\psi_{(n_e,n_d)}^{K'}=\left( \begin{array}{c}
\phi_{(n_e-1,n_d)}^{A}\\
\pm i\phi_{(n_e,n_d)}^{B}\\
\end{array}\right)~.
\ee

We also included the eigenstates for the Dirac point $K'$ which
can be obtained using the same method and display a dispersion
relation identical to the one given by Eq. (\ref{eq1.10}).

\subsection{The Noncommutative Case}

In this section we consider a phase-space noncommutative algebra
as in Eq. (\ref{Eq0.1}). In order to relate the noncommutative
variables $(x_i,p_i)$ with the commutative ones $(x'_i,p'_i)$ we
use the following SW map \cite{CS, OB} (see also \cite{OB2})
\ba\label{eq2.1}
x=x'-{\theta\over 2\hbar}p'_y\hspace{0,2cm},\hspace{0,2cm}p_x=p'_x+{\eta \over2\hbar}y'~,\nonumber\\
y=y'+{\theta\over 2\hbar}p'_x\hspace{0,2cm},\hspace{0,2cm}p_y=p'_y-{\eta \over2\hbar}x'~,\nonumber\\
\ea where $\theta$ and $\eta$ are real constant parameters. The
noncommutative variables then satisfy the algebra
\be\label{eq2.1a}
[x_i,x_j]=i\theta_{ij}\hspace{0,2cm},\hspace{0,2cm}[p_i,p_j]=i\eta_{ij}\hspace{0,2cm},\hspace{0,2cm}[x_i,p_j]=i\hbar_{eff}\delta_{ij}
=i\hbar\delta_{ij}\left(1+{\theta \eta\over4\hbar^2}\right)~, \ee
$i,j=1,2$. This is the noncommutative algebra Eq.(\ref{Eq0.1} with
an effective Planck constant \cite{OB,OB2}. It reduces to
Eq.(\ref{Eq0.1} exactly when $\theta=0$, this being the case we
are going to consider.

Using the same potential vector as in section IIA, Eq.
(\ref{eq1.2}), and substituting the noncommutative variables by
the commutative ones, through the SW map, Eq. (\ref{eq2.1}), we
get,
\begin{displaymath}\label{eq2.2}
(p-eA)^{NC}=\left( \begin{array}{c}
p_x+{eB\over2}y\\
p_y-{eB\over2}x\\
\end{array}\right)=\left( \begin{array}{c}
\lambda p'_x+{eB\over2}\mu y'\\
\lambda p'_y-{eB\over2}\mu x'\\
\end{array}\right)
\end{displaymath}
where
\be\label{eq2.4}
\lambda=\left(1+{{eB\theta}\over{4\hbar}}\right)\hspace{0,2cm},\hspace{0,2cm}\mu=\left(1+{\eta\over{eB\hbar}}\right)~.
\ee
Thus, for the Dirac point $K$, one gets the following Hamiltonian:
\be\label{eq2.5}
H_K=v_F\left( \begin{array}{cc}
0 & \lambda (p_x'-ip_y') +{eB\over2}\mu (y'+ix')\\
\lambda (p_x'+ip_y') +{eB\over2}\mu (y'-ix') & 0
\end{array}\right)~.
\ee

A straightforward comparison with Hamiltonian Eq. (\ref{eq1.3}) shows that noncommutativity reveals itself through constants $\lambda$ and $\mu$. For future convenience one introduces the constant
\be\label{eq2.6}
\gamma=\sqrt{{2\hbar\over eB}{\lambda \over \mu}}=l_B\sqrt{{2\lambda \over \mu}}~.
\ee
In what follows we shall consider only momenta noncommutativity, since in the general Dirac problem, configuration space noncommutativity leads to the breaking of gauge symmetry \cite{Queiroz}, a symmetry preserved in the graphene lattice \cite{Vozmediano}. Indeed, if one evaluates the velocity of a charged particle,
\be\label{eq2.6a}
{\mathbf v}={i\over\hbar}[H_{NC},{\mathbf r}]~,
\ee
we obtain an extra term depending on the $\theta$ parameter, and not the expected result ${\mathbf v}=v_F{ \mathbf\sigma}$. Hence $\lambda=1$ and
\be\label{eq2.6b}
\gamma=\sqrt{{2\hbar\over eB}{1\over\mu}}=l_B\sqrt{{2\over\mu}}~.
\ee
Thus, the equations to be solved are now,
\ba\label{eq2.7}
&&{\hbar v_F\over \gamma}\left[-i\left(\gamma\partial_x-{x\over \gamma}\right)+\left(- \gamma\partial_y+{y\over \gamma}\right)\right]\phi^B=E_K\phi^A~,\nonumber\\
&&{\hbar v_F\over \gamma}\left[-i\left(\gamma\partial_x+{x\over \gamma}\right)+\left(\gamma\partial_y+{y\over \gamma}\right)\right]\phi^A=E_K\phi^B~,
\ea
Following the strategy discussed in the last section, one obtains for the energy spectrum
\ba\label{eq2.8}
E_K^{NC}&=&\pm2\hbar v_F\sqrt{\left({eB\over2\hbar}\right)\left(1+{\eta\over{eB\hbar}}\right)n_e}\nonumber\\
&=&\pm{\hbar v_F\over
l_B}\sqrt{2\left(1+\eta{l_B^2\over{\hbar^2}}\right)n_e}~, \ea
where $n_e$ is a non-negative integer. We clearly see that the
energy at the Dirac point $K$ depends explicitly on the
noncommutative parameter associated with the momenta. Moreover,
one concludes that the noncommutative effect is coupled with the
magnetic field. Of course, the same dispersion relation for energy
is found to the other Dirac point, $K'$. Furthermore, the
eigenvectors for the two Dirac points can be evaluated. One has
\be\label{eq2.9} \psi_{(n_e,n_d)}^K=\left( \begin{array}{c}
\phi_{(n_e,n_d)}^{A}\\
\pm i\phi_{(n_e-1,n_d)}^{B}\\
\end{array}\right)\hspace{0,2cm},\hspace{0,2cm}
\psi_{(n_e,n_d)}^{K'}=\left( \begin{array}{c}
\phi_{(n_e-1,n_d)}^{A}\\
\pm i\phi_{(n_e,n_d)}^{B}\\
\end{array}\right)~.
\ee The eigenvectors can be written in polar coordinates
\cite{Cohen} \be\label{eq2.10}
\phi_{(n_e,n_d)}^L=\mathcal{F}(\rho) e^{im \varphi} e^{-\sigma
\rho^2}~, \ee where $L=A,B$, $\sigma>0$ is some constant,
$\rho^2=x^2+y^2$, $\mathcal{F}(\rho)$ is some polynomial of $\rho$
and $m$ is the quantum number associated to the angular momentum
$l_z$, which is the integer $(m=n_d-n_e)$.

Clearly, the zero-energy level for this system is
$E_K^{NC}=E_{K'}^{NC}=0$. However, the eigenvectors for the
fundamental level are \be\label{eq2.11} \psi_{(0,n_d)}^K=\left(
\begin{array}{c}
\phi_{(0,n_d)}^A\\
0\\
\end{array}\right)\hspace{0,2cm},\hspace{0,2cm}
\psi_{(0,n_d)}^{K'}=\left( \begin{array}{c}
0\\
\pm i \phi_{(0,n_d)}^B
\end{array}\right)~,
\ee meaning that the zero energy for the two Dirac points are
associated with two linearly independent electron states.

\subsection{Quantum Hall effect and noncommutativity}

We conclude from the discussion in the previous section that an
external magnetic field perpendicular to a graphene sheet renders
the energy spectrum discrete. It is known that a charged particle
subjected to an electromagnetic field has a discrete energy
spectrum, the Landau quantization. However, graphene is a special
case, since it is actually a Dirac problem instead of a
Schr\"{o}dinger one \cite{Peres}. The difference between each
Landau level is considerable. This difference in the energy levels
has implications for the quantum Hall effect (QHE).

The QHE is observable in two-dimensional metals, as for instance in bound low-temperature surfaces where electrons are constrained to two dimensions \cite{Neto}. That occurs when the temperature is drastically reduced, the Hall resistivity becomes independent of the magnetic field, and a quantized Hall plateau is formed. Experimentally, the Hall conductivity is given by
\be\label{eq2.12}
\sigma_{xy}={e^2\over h}\nu~,
\ee
where $\nu$ is an integer number and one has the Integer quantum Hall effect (IQHE). However, the QHE observed in graphene is anomalous (AQHE) \cite{Gusynin,Novoselov2}. The difference between the IQHE and the AQHE lies in the fundamental energy level, i.e for the IQHE the first Landau level is observable at zero energy, but for AQHE the first Hall plateau appears when the lowest Landau level is half filled, and the conductivity takes the form,
\be\label{eq2.13}
\sigma_{xy}=\pm4\left(\nu+{1\over2}\right){e^2\over h}~,
\ee
where the factor $4(\nu+1/2)$ is evaluated by taking in account the presence of a zero mode shared by two Dirac points, an that there are $4(\nu+1/2)$ occupied states that are transferred from one edge to another. Notice that this effect shows up at room temperature, contrasting with the usual QHE in semiconductors which is typically a low temperature effect, \cite{Neto,Peres}.

Following Ref. \cite{Streda}, whenever the Fermi level lies in a gap, the Hall conductivity is given by
\be\label{eq2.14}
 \sigma_{H}=e{\partial n(\epsilon_F)\over \partial B}~,
\ee
where $n(\epsilon_F)$ is the density of states, and $\epsilon_F$ is the energy levels of graphene given by Eq. (\ref{eq7}). Thus, in the usual commutative case, the energy levels are given by Eq. (\ref{eq1.10}) and so the density of states is
\be\label{eq2.15}
\epsilon_F= \pm~\hbar v_F \sqrt {2{eB\over\hbar}n_e}\Rightarrow n(\epsilon_F)={eB\over h}~.
\ee
The same strategy can be used to evaluate the Hall conductivity for the noncommutative case. In this case, the energy levels are given by Eq. (36), where the noncommutative correction is explicit. In what concerns the density of states, it seems rather logical that it is affected by the noncommutativity and the simplest way to incorporate this dependence is through the expression:
\be\label{eq2.16}
n(\epsilon_F)={eB\over h}\left(1+{\eta\over{eBh}}\right)~.
\ee
 Thus,
\be\label{eq2.17}
\sigma_H=g_{\epsilon}e\left({e\over h}\left(1+{\eta\over{eBh}}\right)-{eB\over h}{{\eta eh}\over{(eBh)^2}}\right)=g_{\epsilon}{e^2\over h}~,
\ee
where $g_{\epsilon}=4(\nu+1/2)$ is the degeneracy factor. One concludes that even though the momenta noncommutativity induces a change in the energy levels of the graphene electrons, it does not affect the Hall conductivity.

\section{Comparison with experimental results}

In this section, a bound on the noncommutative parameter, $\eta$, can be obtained using the available experimental results. In Ref. \cite{Jiang}, infrared (IR) spectroscopy  in the presence of  a magnetic field was used to resolve the levels of the Landau spectrum for one single layer graphene. Two resonances were resolved for magnetic fields up to $B=18~T$ \footnote{Notice that for magnetic field above $10~T$, the Zeeman energy $g\mu_{B}B$ is negligible in comparison with the energy of the Landau levels \cite{Neto}.}, and their energy position was shown to scale as $\sqrt{B}$ with a slope corresponding to a $v_{F}= (1.12\pm0.02)\times10^{6}~ms^{-1}$ for a particular energy \cite{Jiang}. This value for the Fermi velocity is related with the transition from $n=-1$ to $n=0$ (in the case of holes) and $n=0$ to $n=1$ (in the case of electrons). These transitions correspond to the filling factor $\nu=-2$ in the IQHE, the lowest Landau level transitions possible in graphene.

The energy for the Landau levels $n=1$ or $n=-1$, \be\label{eq3.1}
E_K=\pm\sqrt{2}{\hbar v_{F}\over l_{B}}=\pm\sqrt{2e\hbar v_{F}^2
B}~, \ee where $+$ is for electrons and $n=1$, and $-$ is for
holes and $n=-1$. For these levels one has
$E_K=\pm(172\pm3)~meV$ \cite{Jiang}. Thus, if the noncommutative
energy spectrum for graphene is given by Eq. (\ref{eq2.8}), and
considering that the uncertainty in the energy is at most $6~meV$,
one can obtain a bound for the noncommutative parameter $\eta$. Using Eq. (\ref{eq2.8}) it follows that:
\be\label{eq3.2a}
\eta{l_B^2\over\hbar^2}<0.069~.
 \ee
Thus, the noncommutative parameter $\eta$ satisfies
\ba\label{eq3.3}
&&\eta<2.1\times10^{-53}~kg^2m^2s^{-2}\nonumber\\
&&\Rightarrow\sqrt{\eta}<8.6~eV/c~.
\ea

Naturally,  this bound is not as stringent as the one arising from
the hyperfine transition in the hydrogen atom, $\sqrt{\eta}\leq
2.26~\mu eV/c$, \cite{Queiroz}, one of the most accurate
experimental results in the whole of physics. Despite of that, the
above reasonings show that noncommutative effects are consistent
with what is known about graphene physics.

In what concerns other bounds for the momentum noncommutative
parameter, notice that the one arising from the gravitational
quantum well \cite{OB} and from the equivalence principle
\cite{BOBDP11} depend on an assumption about the configuration
space noncommutative parameter, $\theta$, and cannot the compared
with the above bound without fixing a value for $\theta$.

\section{Conclusions}

In this work the phase-space noncommutative extension of the
graphene in the presence of an external constant magnetic field
was examined. More precisely, only momenta noncommutativity was
considered since the noncommutativity associated with the
configuration variables implies the breaking of gauge invariance.
The introduction of momenta noncommutativity determines a
correction of the energy spectrum of graphene in the presence of a
magnetic field. Moreover, it was shown that this noncommutativity
does not affect the graphene anomalous quantum Hall effect.

Finally, comparison with experimental data reveals that
$\sqrt{\eta}\leq 8.6~eV/c$, a bound that is not very stringent,
but that indicates that there is no contradiction between
noncommutative effects and graphene's physics. These results show
that momentum noncommutativity yields interesting results also at
low-energies and that its implications are not restricted to
quantum comology \cite{Bastos2} and black holes physics
\cite{Bastos3,Bastos4, Bastos5}.

\begin{acknowledgements}

\noindent
The work of CB is supported by Funda\c{c}\~{a}o para a
Ci\^{e}ncia e a Tecnologia (FCT) under the grant SFRH/BPD/62861/2009. The work of OB is partially supported by the FCT project PTDC/FIS/111362/2009. N.C. Dias and J.N. Prata have been supported by the FCT grant PTDC/MAT/099880/2008.

\end{acknowledgements}


\begin{thebibliography}{99}

\bibitem{Novoselov1} K.S. Novoselov et al., Science {\bf 306} (2004) 666; PNAS {\bf 102} (2005) 10451.

\bibitem{Neto} A.H. Castro Neto, F. Guinea, N.M.R. Peres, K.S. Novoselov and A.K. Geim, Rev. Mod. Phys. {\bf81} (2009) 109.

\bibitem{Peres} N. Peres, Rev.Mod.Phys. {\bf 82} (2010) 2673.

\bibitem{Geim} A.K. Geim and K.S. Novoselov, Nature Materials {\bf 6 (3)} (2007) 183.

\bibitem{Gusynin} V.P. Gusynin, and S. G. Sharapov, Phys. Rev. Lett. {\bf 95} (2005) 146801

\bibitem{Novoselov2} K. Novoselov, E. McCann, S. Morozov, V. FalÕko, M. Katsnelson, U. Zeitler, D. Jiang, F. Schedin, and A. Geim, Nature Physics {\bf 2 (3)} (2006) 177.

\bibitem{CS} A. Connes, M.R. Douglas and A. Schwarz, JHEP {\bf 9802} (1998) 003; N. Seiberg and E. Witten, JHEP {\bf 9909} (1999) 032.

\bibitem{DS} M.R. Douglas, N.A. Nekrasov, Rev. Mod. Phys. {\bf 73} (2001) 977; R. Szabo, Phys. Rep. {\bf 378} 207 (2003) 207.

\bibitem{OBLG} O. Bertolami and L. Guisado, JHEP {\bf } (2003) .

\bibitem{OBCZ} O. Bertolami and C. Zarro, Phys. Lett. B {\bf 673} (2009) 83.

\bibitem{Queiroz} O. Bertolami and R. Queiroz, Phys. Lett. A {\bf 375} (2011) 4116.

\bibitem{Nair}  V. P. Nair, A.P. Polychronakros, Phys. Lett. {\bf B 505} (2001) 267.

\bibitem{Gamboa} J. Gamboa, M. Loewe, J.C. Rojas, Phys. Rev. {\bf D 64} (2001) 067901.

\bibitem{DK} M. Demetrian, D. Kochan, Acta Phys. Slov. {\bf 52} (2002) 1.

\bibitem{Horwathy} P.A. Horwathy, Ann. Phys. {\bf 299} (2002) 128.

\bibitem{ZA} Jian-zu Zhang, Phys. Rev. Lett. \textbf{93} (2004) 043002; Phys. Lett. \textbf{B 584} (2004) 204.

\bibitem{OB} O. Bertolami, J. G. Rosa, C. Arag\~ao, P. Castorina and D. Zappal\`a, Phys. Rev. {\bf D 72} (2005) 025010.

\bibitem{OB2} O. Bertolami, J. G. Rosa, C. Arag\~ao, P. Castorina and D. Zappal\`a, Mod. Phys. Lett. {\bf A 21} (2006) 795.

\bibitem{Bastos1} C. Bastos, O. Bertolami, N.C. Dias and J.N. Prata, J. Math. Phys. {\bf 49} (2008) 072101; C. Bastos, N.C. Dias and J.N. Prata, Commun. Math. Phys. {\bf 299} (2010) 709.

\bibitem{Dias} N.C. Dias, M. de Gosson, F. Luef, J.N. Prata, J. Math. Phys. {\bf 51} (2010) 072101.

\bibitem{Bastos9} C. Bastos, O. Bertolami, N.C. Dias and J.N. Prata, Int. J. Mod. Phys. A {\bf 24} (2009) 2741.

\bibitem{Bastos2} C. Bastos, O. Bertolami, N.C. Dias and J.N. Prata, Phys. Rev. {\bf D 78} (2008) 023516.

\bibitem{Bastos3} C. Bastos, O.Bertolami, N.C. Dias and J.N. Prata, Phys. Rev. {\bf D 80} (2009) 124038.

\bibitem{Bastos4} C. Bastos, O. Bertolami, N.C. Dias and J.N. Prata, Phys. Rev. {\bf D 82} (2010) 041502.

\bibitem{Bastos5} C. Bastos, O. Bertolami, N.C. Dias and J.N. Prata, Phys. Rev. {\bf D 84} (2011) 024005.

\bibitem{Falomir1} H. Falomir, J. Gamboa, M. Loewe, F. Mendez and J.C. Rojas, Phys. Rev. {\bf D 85} (2012) 025009.

\bibitem{Falomir2} H. Falomir, J. Gamboa, M. Loewe and M. Nieto, J. Phys. {\bf A45} (2012) 135308.

\bibitem{Cohen} C. Cohen-Tannoudji, B. Diu and F. Lalo\"e, {\it M\'echanique Quantique} (1973) Collection Ensignement.

\bibitem{Vozmediano} M. Vozmediano, M.I. Katsnelson and F. Guinea, Phys. Rep. {\bf 496} (2010) 109.

\bibitem{Streda} P. Streda, J. Phys. {\bf C 15} (1982) L717.

\bibitem{Jiang} Z. Jiang, E.A. Henriksen, L.C. Tung, Y.-J. Wang, M.E. Schwartz, M.Y. Han, P. Kim and H.L. Stormer, Phys. Rev. Lett. {\bf 98} (2007) 197403.

\bibitem{BOBDP11} C. Bastos, O. Bertolami, N.C. Dias and J.N. Prata, Class. Quantum Gravity,  {\bf 28} (2011) 125007.


\end{thebibliography}
\end{document}